\documentstyle[12pt]{article}
\begin{document}
\pagestyle{empty}
\begin{flushright}
{BROWN-HET-896} \\
{February 1993}
\end{flushright}
\vspace*{5mm}
\begin{center}
{\bf Gamma Ray Bursts From Ordinary Cosmic Strings} \\
[10mm]
R.H. Brandenberger, A.T. Sornborger, M. Trodden \\
[10mm]
Department of Physics \\
Brown University \\
Providence RI. 02912. \\
USA. \\[2cm]

{\bf Abstract}
\end{center}
\vspace*{3mm}
We give an upper estimate for the number of gamma ray bursts from ordinary
(non-superconducting) cosmic strings expected to be observed at terrestrial
detectors. Assuming that cusp annihilation is the mechanism responsible for the
bursts we consider strings arising at a GUT phase transition and compare our
estimate with the recent BATSE results. Further we give a lower limit for the
effective area of future detectors designed to detect the cosmic string induced
flux of gamma ray bursts. \\

\newpage\setcounter{page}{1}\pagestyle{plain}

\section{Introduction}
This work is an attempt to extend recent efforts to propose methods to detect
direct evidence for the existence of cosmic strings\cite{M&B 90},\cite{M&B 92}.
Since, after the COBE results\cite{GS& 92},the cosmic string scenario for large
scale structure continues to be in agreement with observation\cite{BSB
92},\cite{LP 93}, we may hope that such methods can provide us with independent
verification (or rejection) of the theory. This will certainly be neccessary if
we are to trust our theoretical predictions. This work concerns only the
behaviour of so called `ordinary' cosmic strings, ie. non-superconducting
strings. Superconducting cosmic strings\cite{WIT 85} have additional phenomena
associated with them\cite{SPG 87},\cite{OTW 86} with which we do not concern
ourselves here.
Motivated by the recent BATSE results from the Gamma Ray Observatory (GRO)
satellite\cite{SF 92} and the need for suggestions for further experimental
techniques, we consider a cosmic string based mechanism with the potential to
account for the observed data and give an upper estimate for the number of
gamma ray burst events which we may expect to observe at a terrestrial detector
due to cosmic strings.

There are several reasons for believing that cosmic strings may be responsible
for gamma ray burst events. Firstly there is the isotropic distribution of the
data which seems to imply an extragalactic origin for the bursters. Any source
at these distances, therefore, needs to have the power to produce gamma rays of
extraordinarily high energy in order that they be observable by the time that
they reach the Earth. Cosmic string mechanisms could naturally account for this
due to the isotropic distribution of the string network and the combined energy
of the mass per unit length of the string ($\sim 10^{15}$ tonnes/cm) coupled
with its relativistic motion. We may also consider the timescale of the bursts;
the mechanism which we investigate, cusp annihilation, may be expected to
produce short, highly directional pulses of radiation in agreement with the
burst observations so far. Further, the spectrum and precise form of the bursts
appears to vary considerably from event to event with no common features. Cusp
an
nihilations might be expected to produce such signatures since different cusps
may be formed on strings of different radii and the extent to which we
intercept a beam from such an event would cause the length of the burst to
vary. Both these effects could give rise to results taking a continuum of
values in a random manner.

Although there are a number of methods by which a cosmic string may produce
high energy gamma radiation we consider only contributions from cusp
annihilations from oscillating string loops\cite{RHB 87},\cite{B&M 87}. We
neglect contributions from cusps formed at string intersections due to
intercommuting and also ignore high energy bursts from string loops emitting
their last energy as particles at the end of their lifetimes. In addition we
neglect the contribution due to particle-antiparticle pairs produced in the
background of a moving string loop. These effects are expected to be small
compared with the dominant contribution from cusp annihilation.

These considerations aside, we consider our result to be an upper estimate
because of our main assumption that at the formation of a cusp all the energy
in the cusp region is released as a burst of high energy particles. In reality
we expect that when the Nambu action breaks down as a cusp is being formed,
back-reaction effects may become important and lead to the release of only a
fraction of the total cusp energy.

\section{Cusp Formation and Annihilation}
The Nambu action for a string of infinitesimal width leads to time periodic
solutions containing at least one cusp per oscillation; ie. parametrizing the
string trajectories as $\underline{x} (s,\tau)$ where $\tau$ is coordinate time
and $s$ parameterizes the length along the string we expect the periodic
formation of points at which $|\underline{\dot{x}}|=1$ and
$\underline{x'}=0$\cite{K&T 82}. As these points begin to form we expect the
Nambu action to break down and strong microphysical forces to counteract the
cusp formation by particle emission.

Initially the energy released by the string will be in the form of false vacuum
quanta of the gauge fields associated with the symmetry breaking giving rise to
the string network. In this work we primarily concentrate on strings produced
at GUT symmetry breakings since it is these that may be responsible for
structure formation. These products will then decay into lower mass particles
and at some stage to quanta to which we may apply the empirical QCD
multiplicity functions to predict the energy spectrum of the final decay
products of the initial particles. In this way we arrive at an expression for
the number of photons expected from cusp annihilation on a string of a given
radius at a given time after the symmetry breaking.

A potentially important point about the emission of this radiation is that it
is highly anisotropic in the rest frame of the loop. In fact the radiation is
beamed into a specific solid angle which we include in our calculation of the
flux expected at terrestrial detectors.

Defining $n_{burst}$ to be the number of observable bursts per unit time we
therefore have the expression

\begin{equation}
n_{burst} = \int_{t_{min}}^{t_{0}} dt \int_{t_{min}}^{t} dR \ \eta(R,t) \
            \frac{1}{R} \ 4\pi d_{c}^{2}(t)
\end{equation}
where $\eta(R,t)$ is the number of strings of radius $R$ per unit volume at
time t, (for this we assume the known scaling solution\cite{AT 85}), $1/R$
gives the rate of cusp formation, $4\pi d_{c}^{2}(t)$ is the comoving area of
the past light cone at time $t$ and $t_{min}$ is the lower cut-off of radius
and time calculated below and is constrained by our detection capabilities.

\section{Detection Considerations}
The number of photons of energy $E$ radiated per unit area at a cusp by a loop
of cosmic string of radius $R$ is given by\cite{M&B 92}

\begin{equation}
N_{R}(E)=\frac{1}{\theta^{2} d^{2}} \frac{\mu l_{c} c^{2}}{Q_{f}^{2}}
         \left[\frac{16}{3} - 2\left(\frac{E}{Q_{f}}\right)^{1/2} -
         4\left(\frac{E}{Q_{f}}\right)^{-1/2} +
         \frac{2}{3} \left(\frac{E}{Q_{f}}\right)^{-3/2} \right]
\end{equation}
Here $\theta^{2}$ is the solid angle into which the radiation is beamed, $d$ is
the physical distance of the loop from the Earth, $Q_{f}$ is the fixed energy
of the particles initially emitted by the cusp (we assume a $\delta$-function
distribution of initial energies) to which we may apply the QCD multiplicity
functions and $l_{c}$ is the length of the overlap region of the loop at the
cusp. Here, as always, $\mu$ represents the mass per unit length of the string.

Now, it is only possible for a detector to register an event as a burst if
sufficient photons are received to distinguish the event from the background of
photons in which the detector operates. If we need $n_{0}$ photons detected to
give a positive detection then we need the corresponding burst at time $t$ to
produce at least $n(t)$ photons, given by

\begin{equation}
n(t)=\frac{4\pi d^{2}(t) n_{0}}{A}\theta^{2}
\end{equation}
where $A$ is the effective area of our detector.
A second constraint on the number of bursts resulting in a detection is the
sensitivity range of the detector. Suppose our detector has a range of
sensitivity of $(E_{min}^{0}, E_{max}^{0})$. Then we may only detect photons
which, after being redshifted on their way to us, have energy lying in this
range. Thus, given a burst by a cosmic string of radius $R$, we require

\begin{equation}
\int_{E_{min}^{0}(1+z(t))}^{E_{max}^{0}(1+z(t))} dE\ N_{R}(E) > n_{0}
\end{equation}
in order that the burst be registered at us (where $z(t)$ is the redshift).

\section{Calculation}
If, as is natural at first, we consider GUT strings then we expect
$Q_{f} \sim 10^{15}$GeV. Clearly, for a typical burst $E << Q_{f}$ and so we
may approximate (2) by

\begin{equation}
N_{R} = \frac{2\mu l_{c}}{3\theta^{2} d^{2}(t) Q_{f}^{2}}
         \left(\frac{E}{Q_{f}}\right)^{-3/2}
\end{equation}
We may then perform the integral (4). Using convenient units and noting that
the overlap length $l_{c} \sim w^{1/3}R^{2/3}$\cite{SPG 87},\cite{RHB 87} where
$w \sim \mu^{-1/2}$ is the width of the string we obtain the inequality

\begin{equation}
\left(\frac{R}{t_{eq}}\right)^{2/3} >
      \left(\frac{E^{0}_{min}}{Q_{f}}\right)^{1/2} (1+z)^{1/2}(t)
      \left(\frac{Q_{f}}{\mu^{1/2}}\right) \left(\frac{d(t)}{t_{eq}}\right)^{2}
      \left(\frac{w}{t_{eq}}\right)^{2/3} \left(\frac{t_{eq}^{2}}{A}\right)
\end{equation}
where we have used the value $n_{0}=10$ for present detectors.
Now, setting $R(t) = t$ in this inequality gives the contribution from the
biggest strings at time $t$ and so provides a lower limit on $t$. This gives

\begin{equation}
\left(\frac{d(t)}{t_{eq}}\right)^{2} \left(\frac{t}{t_{eq}}\right)^{-2/3}
     (1+z)^{1/2}(t) < \left(\frac{E^{0}_{min}}{Q_{f}}\right)^{-1/2}
     \left(\frac{Q_{f}}{\mu^{1/2}}\right)^{-1}
     \left(\frac{w}{t_{eq}}\right)^{-2/3}
\left(\frac{t_{eq}^{2}}{A}\right)^{-1}
\end{equation}

Now, with $E_{min}^{0}$ of order a few MeV and a detector area of 1$m^{2}$ (ie.
the energy range and area of the GRO satellite detector) we have the following
relations
\begin{equation}
\left(\frac{Q_{f}}{E^{0}_{min}}\right) \sim 10^{18}  \ \ \  , \ \ \
 \left(\frac{Q_{f}}{\mu^{1/2}}\right) \sim O(1)
\end{equation}
\begin{equation}
 \left(\frac{A}{t_{eq}^{2}}\right) \sim 10^{-40}  \ \ \ , \ \ \
 \left(\frac{t_{eq}}{w}\right) \sim 10^{51}
\end{equation}
which give us

\begin{equation}
\left(\frac{d(t)}{t_{eq}}\right)^{2} \left(\frac{t}{t_{eq}}\right)^{-2/3}
(1+z)^{1/2}(t) < 10^{3}
\end{equation}
Since it is elementary to show that

\begin{equation}
d(t) = 3t^{2/3}(t_{0}^{1/3} - t^{1/3})
\end{equation}
and we know that $(1+z(t)) = \left(\frac{t}{t_{0}}\right)^{1/2}$ we arrive at

\begin{equation}
\left(\frac{t_{0}}{t}\right)^{1/3}
      \left[\left(\frac{t_{0}}{t}\right)^{1/3} -1 \right]^{2} < 10^{-5}
\end{equation}
Clearly this is only satisfied for $t \in [t_{min},t_{0}]$ where
$t_{min}=t_{0}(1-\epsilon)$ and $\epsilon <<< 1$. Therefore our expression (1)
for the number of observable bursts per unit time becomes

\begin{equation}
n_{burst} = \Delta t \ \eta(R,t_{0}) \ \frac{1}{R} \  d_{c}(t_{min})
\end{equation}
where $t_{min} \sim t_{0}$. This implies that $n_{burst} <<<1$ and completes
our calculation.

\section{Conclusions}
We have obtained a crude estimate for the expected frequency of gamma ray
bursts from cusp annihilations on ordinary GUT cosmic strings. We consider our
result to be an upper bound and, since our calculation shows the flux to be
negligible, we conclude that cusp annihilations from ordinary GUT cosmic
strings are not responsible for the gamma ray burster events observed by the
GRO satellite.

It is easily seen from equations (8)-(12) that if we consider the effective
area of the detector to be the sole parameter in the calculation we may hope to
detect the flux from GUT strings with a detector of $A \sim 10^{5}m^{2}$
ignoring back reaction effects. It is also clear that ordinary strings produced
at lower energy phase transitions will give rise to an even lower estimate of
the flux.

We have shown, therefore, that cusp annihilation is not a viable method of
detection of cosmic strings with forseeable measurement capabilities. It
remains to see whether there exist other phenomena by which we may
independently confirm or reject the existence of these objects.

\end{document}